\documentclass[reprint, amsmath, amssymb, aps]{revtex4-2}

\usepackage{graphicx}
\usepackage{dcolumn}
\usepackage{bm}
\usepackage{xcolor}

\usepackage[normalem]{ulem}

\begin{document}

\preprint{APS/123-QED}

\title{Optical Response by Time-Varying Plasmonic Nanoparticles}

\author{Miguel Verde$^{1,2}$}
\email{miguel.verde@uam.es}
\author{Paloma A. Huidobro$^{1,2,3}$}
\email{p.arroyo-huidobro@uam.es}
\affiliation{
    \vspace{2 pt}
    \parbox{0.746\textwidth}{
        $^1$Departamento de Física Teórica de la Materia Condensada, Universidad Autónoma de Madrid,} \\
    \parbox{0.181\textwidth}{
        E-28049 Madrid, Spain}\\
    \vspace{2 pt}
    \parbox{0.826\textwidth}{
        $^2$Condensed Matter Physics Center (IFIMAC), Universidad Autónoma de Madrid, E-28049 Madrid, Spain}
    \vspace{2 pt}
    \parbox{0.826\textwidth}{
        $^3$Instituto Nicolás Cabrera (INC), Universidad Autónoma de Madrid, E-28049 Madrid, Spain}
    \vspace{7 pt}
    }

\date{\today}

\begin{abstract}
The temporal modulation of material parameters enables optical amplification within linear media. Here we consider the fundamental building block of plasmonics, a subwavelength metal nanoparticle, and study how temporal modulation alters the optical response of the frequency-dispersive scatterers. We show that modulating in time leads to Floquet replicas of the localized surface plasmon resonance of the nanoparticle, which can result in light amplification. We propose a model based on a point-like dipole description of the time-varying frequency-dispersive nanoparticle that fully captures the radiative and amplifying properties of the system in the subwavelength regime. By comparing our simplified model to full Floquet-Mie scattering calculations, we demonstrate that the optical scattering by the nanoparticle is accurately described by an analytical two-band model. This allows us to introduce a two-frequency effective polarizability that fully incorporates the properties of the localized surface plasmon and its amplifying replica, as well as their interaction. In addition, we analyze the emergence of the parametric amplification condition for the modulated nanoparticle, showing that amplification can be obtained in a broad range of parameters. 
\end{abstract}

\maketitle


\section{\label{sec:level1}Introduction} 

    \begin{figure}[t]
        \includegraphics[width=0.95\columnwidth]{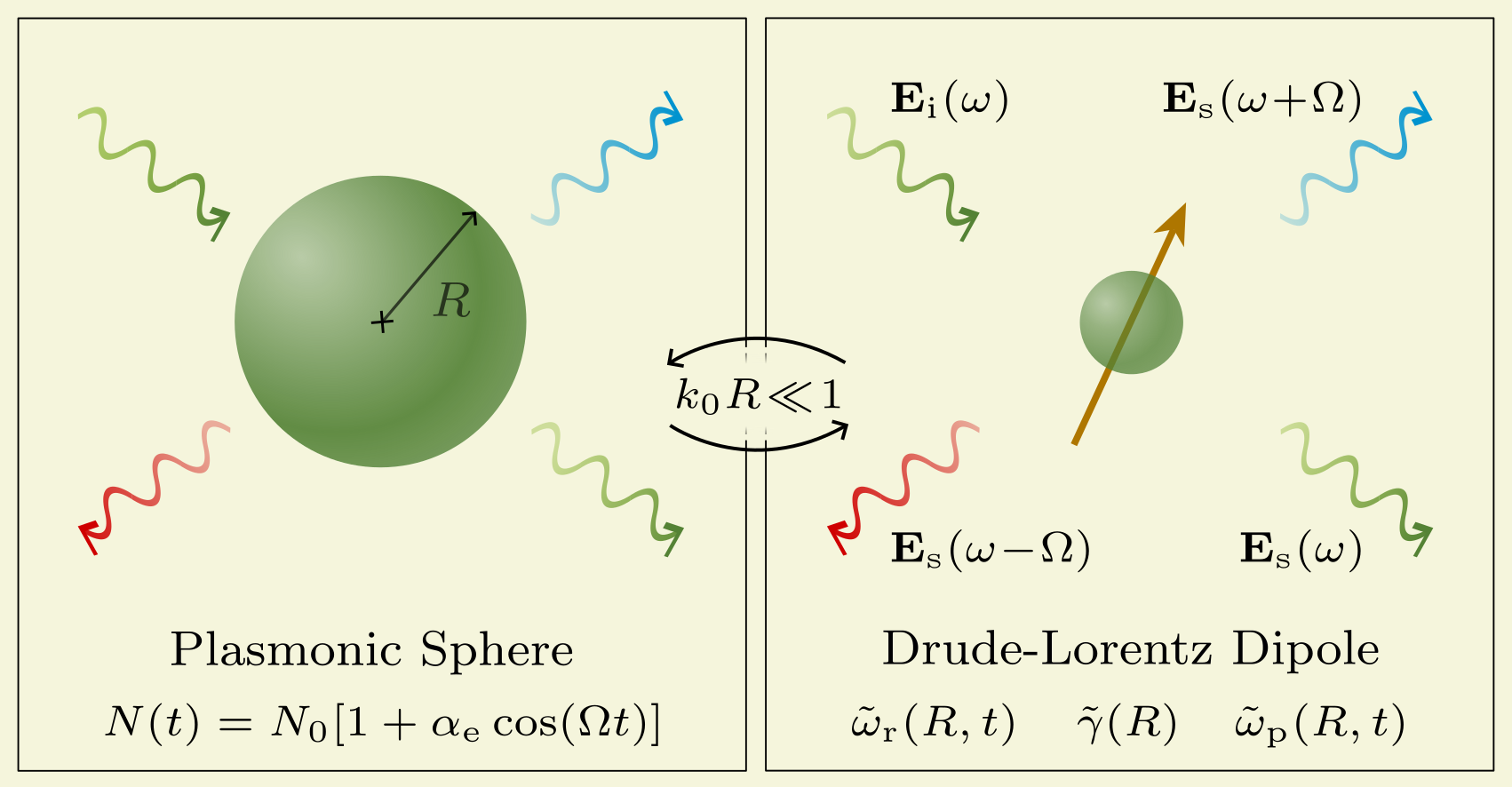}
        \caption{Illustration of the system under study: A plasmonic nanosphere with a carrier density that changes periodically in time, mapped onto a time-varying Drude-Lorentz point dipole within the subwavelength regime.}
    \end{figure}

    \noindent The study of materials with time-dependent optical properties has drawn growing attention due to the possibilities unlocked by loosening the constraints enforced by energy conservation, passivity and reciprocity \cite{review,pacheco-pena-2022, engheta-2023}. Such media have been proposed for diverse wave manipulation applications, from amplification and linear frequency conversion to non-reciprocal photonics \cite{1124533, 1139657, mendonca-2002, yu-2009, sounas, galiffi-2019, zhou-2020}. In this scenario, photonic time crystals (PTCs) have emerged as a subject of particular interest. These are media whose refractive index undergoes ultrafast and large periodic variations in time \cite{martinez-romero-2018, PTC, boltasseva-2024}. Notably, PTCs have momentum band gaps which support complex-valued eigenfrequencies responsi- ble for the amplification of classical fields \cite{lyubarov-2022, gaxiola-luna-2023} and the quantum vacuum \cite{mendonca-2000, ganfornina-andrades-2024, sustaeta-osuna-2025}. 
    
    Temporal modulation of media is possible at different frequency ranges and on various platforms. For instance, in the microwave regime, metasurfaces and transmission lines can implement circuit element modulations through electronic means \cite{wang-prueba, reyes-ayona-2015, zhang-2018, zhao-2018}, while at higher frequencies, low Drude weight semiconductors, such as indium tin oxide and other transparent conducting oxides, support a fast and unusually strong variation of the refractive index in the near-IR \cite{alam-2016, tirole}. These materials, typically described as plasmonic media, i.e. Drude-like media, exhibit strong dispersion in the epsilon-near-zero regime where a large refractive index variation can be realized  \cite{alam-2016, vezzoli-2018, lobet-2023}.

    On the other hand, plasmonic materials are a broadly used platform for applications such as energy harvesting, surface enhanced spectroscopy and sensing \cite{barnes-2003, kuhn-2006, catchpole-2008,garcia-vidal-2010, ross-2015, fernandez-dominguez-2017}. This wide range of applications stems from their enhanced interaction with light, which gives rise to large absorption and scattering cross-sections, as well as notable enhancements of electromagnetic energy in the near-field \cite{maier-2005,maier-2007}. These effects result from collective oscillations of Drude-like, free-charge carriers confined within the conductive material, the well-known surface plasmons. Surface plasmons are tunable through the spatial structuring of the plasmonic medium, and this has been extensively studied in the past \cite{giannini-2010}. However, time-modulated plasmonic nanostructures and the phenomenology deriving from the modulation remains largely unexplored.

    Notably, PTCs have been primarily studied as spatially unbound and dispersionless, time-modulated media, that is, infinite media with instantaneous material response to an external field \cite{lyubarov-2022, gaxiola-luna-2023, mendonca-2000, ganfornina-andrades-2024, sustaeta-osuna-2025, pacheco-pena-2021, galiffi}. In contrast, any physical implementation of a PTC requires modulating a realistic material, that will generally be dispersive and absorbing, and take the form of a finite structure. Hence, a growing number of recent studies focus on formulating theoretical models that integrate modulation in time with both finite structures \cite{stefanou,  globosits-2024, sadafi-2025, valero2025} and dispersion \cite{galiffi-2020, horsley-2023, Floquet-Mie, asadchy, sloan-2024, raziman-2025, allard-2025}. In this context, dispersion has been shown to enable enlarged momentum band gaps beyond the capabilities of dispersionless PTCs \cite{wang-2024}. Moreover, the properties of periodically modulated Mie spheres have been studied, revealing effects induced by the modulation such as directional scattering and light amplification \cite{Floquet-Mie, asadchy}, as well as the collective response of ensembles of time-varying scatterers \cite{garg-2022,sadafi-2025}.

    In this work, we investigate the optical properties of a periodically time-dependent plasmonic nanoparticle, that constitutes the fundamental building block of plasmonics. Specifically, we consider spherical particles in the subwavelength regime, with a time-varying carrier density, and we analyze the physical mechanisms arising from a periodic temporal modulation. We show that such time-varying material's properties enable new resonant modes within the particle, which unlock the possibility of amplification, overcoming material and radiative damping. In addition, we introduce a simplified model for such subwavelength time-dependent plasmonic particles in the form of a parametric Drude-Lorentz electric point-dipole, which allows us to describe analytically the optical response of the nanosphere within a two-mode approximation. Our simplified analytical description fully captures the behavior of time-varying periodic nanoparticles within the subwavelength regime.

    The article is organized as follows: In section II, we introduce the Floquet T-matrix formalism and analyze the scattering properties of time-varying plasmonic particles. In section III, we revisit the electric dipole approximation for subwavelength nanoparticles in the scenario where the material properties are modulated in time, and introduce the two-time polarizability. In section IV, we present an analytical two-band model for a time-varying nanosphere that describes it as a two-frequency Drude-Lorentz dipole and that accurately replicates its optical response. In section V, we utilize the analytical model to elucidate the optical properties of time-dependent nanoparticles derived from full Floquet T-matrix computations. Finally, conclusions are provided in section VI.

\section{\label{sec:level2}Scattering by a Time-Dependent Plasmonic Nanoparticle}

\subsection{T-matrix formalism for time-varying scatterers}
    \noindent We analyze the optical response of time-varying, isotropic and homogeneous plasmonic nanoparticles with spherical geometry. As a first step, we revisit the optical properties of the bulk plasmonic medium assuming a periodic varia-tion of its charge-carrier density:
    \begin{equation}
        N(t)=N_0[1+\alpha_\mathrm{e}\cos(\Omega t)],
    \end{equation}
    where $N_0$ is the average charge-carrier density, $\alpha_\mathrm{e}$ is the modulation strength and $\Omega$ is the modulation frequency. Electromagnetic waves propagating in these media satisfy Maxwell's equations, $\nabla\times\textbf{E}=-\partial_t\textbf{B}$ and $\nabla\times\textbf{H}=\partial_t\textbf{D}$, along with the constitutive relations 
    \begin{subequations} \label{eq:const}
    \begin{align}
    \textbf{D}(\textbf{r},t) &= \epsilon_0\textbf{E}(\textbf{r},t) + \textbf{P}(\textbf{r},t) , \label{eq:const_a} \\
    \textbf{B}(\textbf{r},t) &= \mu_0\textbf{H}(\textbf{r},t) , \label{eq:const_b}
    \end{align}
    \end{subequations}
    with $\textbf{E}$ and $\textbf{H}$ the external electric and magnetic fields, $\textbf{D}$ and $\textbf{B}$ the displacement field and magnetic flux density, $\textbf{P}$ the polarization density, and $\varepsilon_0$ vacuum's permittivity. In particular, the polarization density obeys a Drude-like transport equation,
    \begin{equation}
        \left[ \partial^2_t + \gamma\,\partial_t \right]\textbf{P}(\textbf{r},t) = 
        \varepsilon_0\omega^2_\mathrm{p}(t)\textbf{E}(\textbf{r},t),
    \end{equation}
    where $\gamma$ represents the ohmic losses, and $\omega_\mathrm{p}(t)$ is the time dependent plasma frequency:
    \begin{equation}
        \omega^2_\mathrm{p}(t)=\frac{e^2}{\varepsilon_0 m_\mathrm{e}}N(t)\equiv\omega^2_\mathrm{ps}[1+\alpha_\mathrm{e}\cos(\Omega t)],
    \end{equation}
    with $m_\mathrm{e}$ the effective electron mass within the materials, $e$ the elementary charge and $\omega_\mathrm{ps} $ the plasma frequency in the absence of modulation. Notably, such a description implies that the electric field at any given time interacts only with the immediate charge-carriers population. While alternative formalisms have been suggested to describe the influence of charge-carrier modulation on the polarization densities \cite{stepanov-1976, Polarizability, horsley-2023}, they all share the phenomenology of gain and frequency conversion that we study in this work.

    For a time-varying material driven by the constitutive relations in Eq.$\,$\ref{eq:const}, the polarization density is described by \newline the two-time susceptibility: $\chi_\mathrm{e}(t,t-\tau)$. Such a function characterized the polarization density induced at a time $t$ by an electric field applied at an earlier time $\tau$,
    \begin{equation}
        \textbf{P}(\textbf{r},t)=\varepsilon_0\int d\tau\, \chi_\mathrm{e}(t, t-\tau)\,\textbf{E}(\textbf{r},\tau).
    \end{equation}
    Because of the periodic nature of the medium's properties, the electric susceptibility satisfies the discrete time-translation symmetry relation:
    \begin{equation}
        \chi_\mathrm{e}(t,t-\tau)=\chi_\mathrm{e}(t+n\cdot2\pi/\Omega,t-\tau),
    \end{equation}
    with $n$ an integer number. As a result, its double Fourier transform becomes discrete in its first frequency variable,
    \begin{equation} 
        \overline\chi_\mathrm{e}(\omega_\mathrm{a},\omega_\mathrm{b}) =\sum_{n=-\infty}^\infty\delta(\omega_\mathrm{a} - n\cdot\Omega)\,\overline\chi_e(\omega_\mathrm{a},\omega_\mathrm{b}),
    \end{equation}
    where $\delta(\omega)$ is the Dirac delta function and we adopt the Fourier transform convention as defined in \cite{Floquet-Mie}. The wave equation for the electric field within the bulk medium is then expressed in terms of the following set of equations:
    \begin{multline}
        \nabla \times\nabla\times\overline{\textbf{E}}(\textbf{r},\omega_v)=  k_0^2(\omega_v)\cdot  \\
       \left[ \overline{\textbf{E}}(\textbf{r},\omega_v) + \sum_n \overline{\chi}_e(\omega_v-\omega_n, \omega_n)\overline{\textbf{E}}(\textbf{r},\omega_n)\right],
    \label{eq:wave}
    \end{multline}
    where $\omega_n=\omega +n\,\Omega$ is the $n$-th frequency harmonic, with $\omega\in[0,\Omega)$ the Floquet frequency, and $k_0(\omega)=\omega/c$ is the free space wavenumber. In the plasmonic medium under consideration, the Fourier-space susceptibility is given by
    \begin{multline}
        \overline{\chi}_\mathrm{e}(\omega_v-\omega_n,\omega_n)=\left\lbrace\varepsilon_\mathrm{s}(\omega_v)-1\right\rbrace\delta_{vn} \\
        +\varepsilon_\mathrm{d}(\omega_v) \left\lbrace\delta_{vn-1} + \delta_{vn+1}\right\rbrace,
    \end{multline}
    with $\delta_{nv}$ being the Kronecker delta. The stationary and dynamic relative permittivities are defined as,
    \begin{equation}
        \varepsilon_\mathrm{s}(\omega)=\varepsilon_\infty-\frac{\omega_{ps}^2}{\omega^2+i\omega\gamma},
        \hspace{5pt}
        {\varepsilon}_\mathrm{d}(\omega)=-\frac{\alpha_\mathrm{e}}{2}\frac{\omega_{ps}^2}{\omega^2+i\omega\gamma},
    \end{equation}
    with $\varepsilon_\infty$ denoting the high frequency dielectric constant which accounts for the contribution of bound electrons. The wave equation \ref{eq:wave} is solved numerically by truncating the frequency spectrum to a spectral window of interest. In addition, we use the method of separation of variables and look for solutions that satisfy the ansatz:
    \begin{equation}\label{eq:ansatz}
        \overline{\textbf{E}}(\omega, \textbf{r})=\int d\kappa\,\mathcal{A}(\kappa)S_\kappa(\omega)\textbf{F}_\kappa(\textbf{r}),
    \end{equation}
    where $\mathcal{A}(\kappa)$ is a complex amplitude and $\kappa^2$ represents the separation constant. The spatial and spectral part of the eigenvectors satisfy the following set of coupled equations:
    \begin{subequations} \label{eq:eigen}
    \begin{align}
    \mathcal\nabla \times\nabla\times \textbf{F}_\kappa(\textbf{r}) = \kappa^2 \textbf{F}_\kappa(\textbf{r})&, \label{eq:eigen_a} \\
    S_\kappa(\omega_v) + \sum_l\overline{R}_e(\omega_v-\omega_l, \omega_l)S_\kappa(\omega_l) 
    = \frac{\kappa^2}{k_v} S_\kappa(\omega_v)&, \label{eq:eigen_b}
    \end{align}
    \end{subequations}
    with $k_{v}\equiv k_0(\omega_v)$. In spherical coordinates, the solutions 
    of the spatial equation are given by the vector spherical harmonics $\textbf{F}^{\,\iota}_{\beta,lm}\left(\kappa\textbf{r}\right)$. The index $\iota$ distinguishes between radiating $(\iota=1)$ and regular $(\iota=3)$ spherical harmonics. In addition, $l$ is the multipolar order of the wave vector, and $m$ denotes the angular momentum along the z-axis. Finally, the label $\beta$ indicates the polarization of the wave, which can be either transverse electric \{$M$\} or transverse magnetic \{$N$\} \cite{VSH}. This approach allows us to determine the eigenvector of the bulk material, which are needed for computing the optical properties of the nanoparticle \cite{Floquet-Mie}.

    Now we consider the optical properties of the time-varying spherical nanoparticles embedded in vacuum. To address this scattering problem, we use the Floquet-Mie T-matrix formalism developed by Ptitcyn et al.$\,$\cite{Floquet-Mie}. Using this approach, we evaluate the optical response of the particles through the two-time electric response function. To extract the $T$-matrix of the sphere, we express the scattered and incident fields in terms of vector spherical harmonics,
    \begin{align}
        \overline{\textbf{E}}_{\mathrm{sca}}(\textbf{r},\omega_v) &=\sum\nolimits_{\beta,l,m} E^\mathrm{sca}_{\beta,lm}(\omega_v)\textbf{F}^{(3)}_{\beta,lm}\left(k_v \textbf{r}\right),\\
        \overline{\textbf{E}}_{\text{inc}}(\textbf{r},\omega_v) &=\sum\nolimits_{\beta,l,m} E^\mathrm{inc}_{\beta,lm}(\omega_v)\textbf{F}^{(1)}_{\beta,lm}\left(k_v \textbf{r}\right).
    \end{align}    
    Using the eigenstates of the bulk material given by Eq.$\,$\ref{eq:ansatz}, and prescribing continuity conditions for both transverse electric and magnetic fields at the surface of the particle, the unknown complex amplitudes $E^{sca}_{\beta,lm}$ are derived from the known incident amplitudes. The connection is given by the T-matrix \cite{Floquet-Mie}:
    \begin{equation}
        \vec{E}^{\,\mathrm{sca}}_{\beta,lm}(\omega) = \hat{\text{T}}_{\beta,l}^\mathrm{sca}(\omega)\cdot\vec{E}^{\,\mathrm{inc}}_{\beta,lm}(\omega),
    \end{equation}
    where we define the column vector:
    \begin{equation}
        \vec{E}_\lambda(\omega)=\left[\cdots\;  E_\lambda(\omega_{-1})\;E_\lambda(\omega_0)\; E_\lambda(\omega_1)\;\cdots\right]^T,
    \end{equation}
    with $\lambda$ labeling the degrees of freedom of the electric field. In general, the T-matrix depends on the particle radius, 
    the eigenstates of the time-varying bulk material, and the environment refractive index; but it is independent of the angular momentum projection $m$ along the z-axis.

    We focus our study on the absorption cross-section of time-dependent nanoparticles. This parameter quantifies the rate at which electromagnetic field energy is absorbed by a system embedded in a non-dissipative environment when excited by an external field. For a monochromatic incident plane wave at frequency $\omega_n$,
    \begin{equation}
        \overline{\textbf{E}}_{\text{inc}}(z, t) = \frac{1}{\sqrt{2\pi}}E_0\exp\left[i(\omega_n t - k_n z) \right]\hat{x},
    \end{equation}
    the normalized absorption cross-section can be evaluated at each harmonic frequency $\omega_v$ and is given by:
    \begin{equation}\label{eq:sig_abs}
        \sigma_{\text{abs}}^{[v,n]}(\omega)=\delta_{vn}\,\sigma_{\text{ext}}^{[v,n]}(\omega)-\sigma_{\text{sca}}^{[v,n]}(\omega),
    \end{equation}
    where $\sigma_{\text{sca}}$ and $\sigma_{\text{ext}}$ denote the normalized scattering and extinction cross-sections, respectively \cite{bohren}:
    \begin{gather}
        \sigma^{[v,n]}_{\textrm{sca}}(\omega)=\frac{2\pi}{k_0^2(\omega_v)A }\sum_{\beta,l}(2l+1)\left|\lbrace {\hat{T}^\mathrm{sca}_{\beta,l}(\omega)}\rbrace_v^n\right|^2,\\
        \sigma^{[v,n]}_{\text{ext}}(\omega)=-\frac{2\pi}{k_0^2(\omega_v)A }\sum_{\beta,l}(2l+1)\Re\left[\lbrace {\hat{T}^\mathrm{sca}_{\beta,l}(\omega)}\rbrace_v^n\right],
    \end{gather}
    with $\lbrace\hat{T}^{sca}_{\beta,l}(\omega)\rbrace_v^n$ being a matrix element of the T-matrix, and $A=\pi R^2$ the nanoparticle's area under illumination. 
    

\subsection{Optical spectrum of periodic time-dependent subwavelength nanoparticles}    

    \begin{figure*}[t]
        \includegraphics[width=\textwidth]{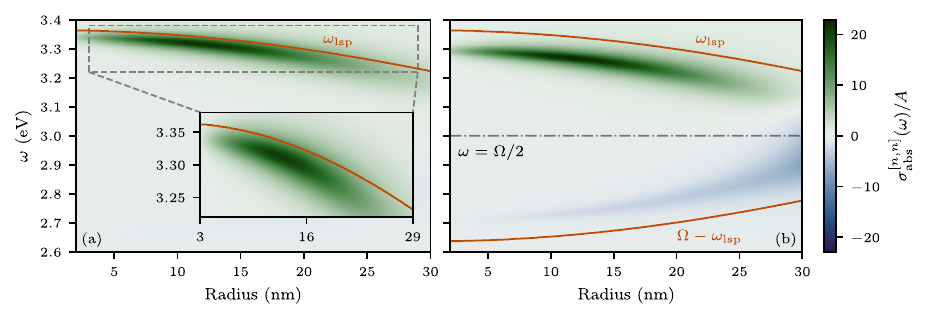}
        \caption{Normalized absorption cross-section of a temporally modulated silver sphere as function of the nanoparticle radius and for two values of the modulation strength: $\alpha_\mathrm{e}=0.15$ (a) and $0.25$ (b). Modulation frequency is fixed at $\Omega=6.0$ eV $=0.67\,\omega_{p,0}$. The inset shows a zoomed-in view of the region enclosed in the square marked by the dashed gray line. The orange lines mark the localized surface plasmon resonance in the absence of temporal modulation, $\omega_{lsp}$, and the first order Floquet replica of its negative frequency copy, $\Omega-\omega_{lsp}$.}
    \end{figure*}

    \noindent   In the subwavelength regime, defined by $x\equiv k_0(\omega) R\ll 1$ for a nanoparticle of radius $R$, it is well known that plasmonic nanoparticles respond to external fields as electric dipoles. We now use the T-matrix formalism introduced in the previous section to analyze the optical response of time-modulated plasmonic nanoparticles. Specifically, we use Eq.$\,$\ref{eq:sig_abs} to compute the absorption spectra of subwavelength plasmonic nanospheres. Notably, in the time-modulated scenario, where different frequencies are coupled, the subwavelength regime is defined by,
    \begin{equation}
        x_n\equiv k_0(\omega_n)R\ll 1.
    \end{equation}
    In addition, throughout this work, the parameters of the plasmonic medium are taken to be those of silver: $\omega_\mathrm{ps}= 8.9$ eV, $\gamma= 38$ meV and $\varepsilon_\infty=5$ \cite{Silver}. In the conventional non-modulated setting, the absorption cross-section of a subwavelength plasmonic nanoparticle exhibits a distinct peak at the localized surface plasmon (LSP) resonant fre- quency, $\omega_\text{lsp}$. For $x \rightarrow 0$, the nanoparticle is resonant at:
    \begin{equation}
        \omega^2_\mathrm{q}=\lim_{x\rightarrow0}
        \omega_\mathrm{lsp}^2(R)=\omega_\mathrm{ps}^2/(\varepsilon_\infty+2).
    \end{equation}
    As the radius of the nanoparticle grows, the maximum of the normalized absorption spectrum also increases, with the LSP resonance redshifting as the radius grows. However, for a certain size (around 12 nm for silver nanoparticles) retardation effects become more important, broadening the resonance and reducing the maximum of the normalized absorption spectrum through radiative losses.

    The effect that a periodic modulation of the material's parameters has on the optical properties of the plasmonic nanoparticle are depicted in Fig.$\;$2. We illustrate density plots of the normalized absorption spectra as a function of the system's size, calculated with the full T-matrix (Eq.$\,$\ref{eq:sig_abs}). We take a fixed value of modulation frequency, $\Omega=6.0$ eV (smaller than silver's plasma frequency) and two values of modulation strength: a weaker modulation, $\alpha_\mathrm{e} = 0.15$ (a), and a stronger modulation,  $\alpha_\mathrm{e} = 0.25$ (b). For the lower modulation strength, the particle exhibits a single resonant mode, identified by a maximum in the absorption spectrum. This resonant eigenmode is the LSP, as in the static system, red-shifted due to the periodic temporal modulation. This can be seen by comparing to the orange line, which signals the LSP dispersion in the absence of modulation. The inset panel shows a zoomed-in view of the resonant mode to improve its visualization. The LSP of the time-varying nanoparticle behaves as the conventional one: it further redshifts as the radius grows and exhibits a maximum in the absorption cross-section around  $R\sim 12\,\text{nm}$. 

    In Fig.$\;$2(b), we show the effects of a stronger temporal modulation with $\alpha_\mathrm{e} = 0.25$. In this case, the LSP behaves similarly to the weakly modulated scenario (a), although with a more pronounced redshift compared to the LSP of the unmodulated nanosphere (plotted again as an orange line). In addition, we find that increasing  the modulation strength leads to the emergence of a new resonance with negative absorption cross-section, seen in blue in the plot. Unlike conventional resonances in plasmonic systems, the new resonance is associated with optical amplification. Notably, it emerges as the first-order Floquet replica of the negative frequency copy of the LSP, such that it follows $\omega=\Omega-\omega_\mathrm{lsp}(R)$ (see orange line), but this time with a blueshift. For this reason, the new resonance's dispersion with the radius is mirror symmetric to that of the LSP with respect to $\omega=\Omega/2$. In addition, and contrary to what we observe in the conventional LSP, the height of this minimum increases monotonically with $R$, and it also broadens, surviving at radii values for which the absorption cross-section of the original LSP becomes very small due to radiative loss.    

\section{\label{sec:level3} Electric Dipole Approximation Revisited For Time-Varying Particles}

    \noindent As we have seen, the optical response of a time-dependent plasmonic particle in the subwavelength regime is dominated by the dipolar electric resonances, the LSP and its negative-frequency counterpart. In the following section we revisit the electric dipole approximation for the time-varying case and derive an equation for the polarizability of the scatterer, which, in contrast to the conventional case, is now a two-frequency polarizability.

    First, within the Floquet-Mie framework, we can write the field scattered by time-varying nanoparticles (Eq.$\;$13) within the subwavelength regime as just the contribution of modes with dipolar electric character: $\lbrace\beta,l\rbrace=\lbrace N,1\rbrace $. Hence, for a given Floquet harmonic $v$, the scattered field reads as,
    \begin{equation}\label{eq:bpar}
        \overline{\textbf{B}}_\mathrm{sca}(\textbf{r},\omega_v) = \frac{3i}{2c}E_0\lbrace\hat{T}^\mathrm{sca}_{N,1} \rbrace_v^n\left[\hat{r}\times\hat{x}\right] h_1^{(1)}(k_v r),
    \end{equation}
    where $h^{(1)}_l(x)$ is the spherical Hankel function of the first kind of order $l$, and we have used the coefficients from the plane wave expansion in terms of regular vector spherical harmonics,
    \begin{equation}
        E^\mathrm{\,inc}_{\beta,lm}(\omega_n)=2\pi i^{l+1}E_0\gamma_{l,-1}\lbrace\delta_{m,1}+(-1)^{\delta_{\beta N}}\delta_{m,-1}\rbrace,
    \end{equation}
    with $\gamma_{l,m}$ being the normalization coefficient of the vector spherical harmonics \cite{VSH}. We note that we have numerically verified that the T-matrix elements satisfy:
    \begin{equation}
        \lbrace {\hat{T}^\mathrm{sca}_{\beta,l}(\omega)}\rbrace_v^n\approx 0 \text{, if } \lbrace\beta,l\rbrace\neq\lbrace N,1\rbrace,
    \end{equation}
    which extends the analytical approach used for Mie-coefficients in subwavelength non-modulated scatterers \cite{bohren}.  

    On the other hand, we now consider a time-modulated electric point dipole. The response of such a scatterer is driven by a two-time dependent polarizability: $\alpha(t,t-\tau)$. Hence, the dipole moment verifies the following equation when excited by an external field \cite{Polarizability}:
    \begin{equation}
        \textbf{p}(\textbf{r},t)=\int d\tau\,\alpha(t,t-\tau)\textbf{E}_\mathrm{inc}(\textbf{r},\tau).
    \end{equation}
   Assuming a time-periodic modulation as in the previous section and Fourier transforming to the spectral domain, we derive the identity:
    \begin{equation}
        \overline{\textbf{p}}(\textbf{r},\omega_v)= \sum_k \overline\alpha(\omega_v-\omega_k,\omega_k) \overline{\textbf{E}}_\mathrm{inc}(\textbf{r},\omega_k).
    \end{equation}
    Scattering by such a point-like dipole upon excitation by a monochromatic plane wave with frequency $\omega_n$, leads to the following magnetic field at a harmonic frequency $\omega_v$:
    \begin{equation}\label{eq:bdip}
        \overline{\textbf{B}}_\mathrm{dip}(\textbf{r},\omega_v) = \frac{1}{c}\frac{k_v^3}{4\pi\varepsilon_0}\left[\,\overline{\textbf{p}}(\omega_v)\times\hat{r}\,\right]h^{(1)}_1(k_{v}r),
    \end{equation}
    with the dipole located at the origin of coordinates, as in the T-matrix formalism for the nanoparticle \cite{Dipole-Scat}.

    We can now compare the expressions for the magnetic fields scattered by the subwavelength plasmonic scatterer (Eq.$\,$\ref{eq:bpar}) and by the time-dependent point dipole (Eq.$\,$\ref{eq:bdip}), thereby deriving an identity that links the elements of the two-time polarizability tensor with the Floquet elements of the T-matrix:
    \begin{equation}\label{eq:conex}
        \overline\alpha(\omega_v-\omega_n,\omega_n)= -\frac{3i}{2k_0^3(\omega_v)} 4\pi\epsilon_0\lbrace\hat{T}^\mathrm{sca}_{N,1}(\omega) \rbrace_v^n\,.
    \end{equation}
    This expression is the generalization of the one that connects the polarizability of a static electric dipole, $\alpha_\mathrm{s}$, with the first-order electric Mie coefficient of a subwavelength conventional sphere, $a_1(\omega)$ \cite{bohren}:
    \begin{equation}
        \overline\alpha_\mathrm{s}(\omega)= 6i\pi\epsilon_0 a_1(\omega)/k_0^3(\omega).
    \end{equation}
    With Eq.$\,$\ref{eq:conex}, we have derived a closed expression for the two-frequency polarizability of a nanosphere that is virtu- ally exact in the subwavelength regime. Finally, we may formulate the normalized scattering and extinction cross-sections of the nanoparticle (or the time-varying electric point dipole) in terms of its dipolar polarizability as:
    \begin{gather}
        \sigma^{[v,n]}_\mathrm{sca,p}(\omega) = \frac{k_v^4}{6\pi\varepsilon_0^2A} \left|\,\overline\alpha(\omega_v-\omega_n,\omega_n)\right|^2, \\
        \sigma^{[v,n]}_\mathrm{ext,p}(\omega) = \frac{k_v}{\varepsilon_0A} \Im\left[\,\overline\alpha(\omega_v-\omega_n,\omega_n)\right].
    \end{gather}
    Both expressions are obtained by substituting Eq.$\,$\ref{eq:conex} into the cross-section formulas of a time-varying nanoparticle (Eqs.$\;$19 and 20). Additionally, the absorption spectrum is derived from the above expressions. 

    \begin{figure*}[t]
        \includegraphics[width=\textwidth]{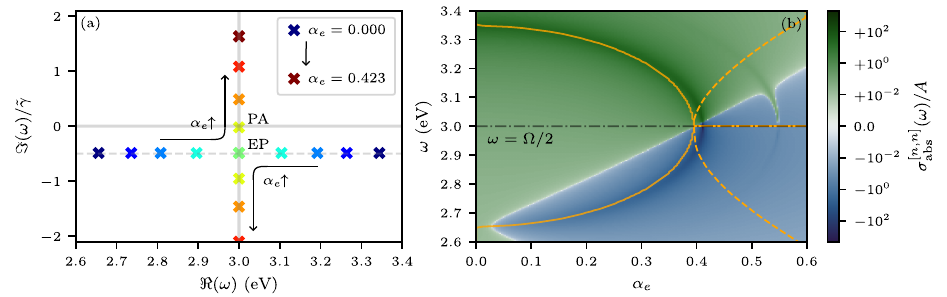}
        \caption{(a) Eigenmodes and (b) normalized absorption cross-section of the time-varying plasmonic particle with radius $R=10$ nm, as a function of the modulation strength $\alpha_\mathrm{e}$, at a fixed $\Omega=6$ eV. The orange solid and dashed lines represent the real and imaginary parts of the system's eigenstates given by the two-band model, respectively.}
    \end{figure*}    

\section{\label{sec:level4}Two-Band Description}

    \noindent After revisiting the electric dipole approximation and deriving a two-frequency polarizability for time-modulated nanospheres, we now propose a Drude-Lorentz point-like dipole model with time-varying coefficients. This description appropriately captures all the ingredients needed to model the scattering by the time-dependent nanoparticle.
    

    In particular, we model the time-dependent plasmonic sphere as an electric point-like dipole whose dynamics are governed by a Drude-Lorentz differential equation:
    \begin{equation}\label{eq:drude-lorentz}
        \left[ \partial^2_t + \tilde\gamma \,\partial_t +\tilde\omega_\mathrm{r}^2(t)\, \right]\textbf{p}_\mathrm{L}(t)=\varepsilon_0\tilde\omega_\mathrm{p}(t)\textbf{E}_\mathrm{inc}(t).
    \end{equation}
    where $\tilde\omega_\mathrm{r}(t)$ denotes the time-varying resonant frequency of the dipole, $\tilde\omega_\mathrm{p}(t)$ is the time-dependent effective plasma frequency, and the damping factor $\tilde\gamma=\gamma+\gamma_\text{rad}$ includes both ohmic and radiative losses, with
    \begin{equation}
        \gamma_{\text{rad}}=2R^3\omega_{\text{lsp}}^6(R)/(c^3\omega_\text{p}^2).
    \end{equation}
    The temporal dependence of the Drude-Lorentz frequencies is 
    taken following the modulation of the plasma frequency in the original model:
    \begin{gather}
        \tilde\omega^2_\mathrm{r}(t)=\tilde\omega^2_\mathrm{rs}[1+\alpha_\mathrm{e}\cos(\Omega t)],\\
        \tilde\omega^2_\mathrm{p}(t)=\tilde\omega^2_\mathrm{ps}[1+\alpha_\mathrm{e}\cos(\Omega t)].
    \end{gather}   
    Here, $\tilde\omega_\mathrm{rs}$ and $\tilde\omega_\mathrm{ps}$ are taken to reproduce the optical properties of a static sphere in the subwavelength regime, such that the resonant frequency is given by the LSP frequency of the nanoparticle, $\tilde\omega_\mathrm{rs} = \omega_\mathrm{lsp}\,$; and the effective plasma frequency relates to the nanoparticle's parameters through:
    \begin{equation}
        \tilde\omega_\mathrm{ps}^2=12\pi R^3\omega_\mathrm{lsp}^4(R)/\omega_\mathrm{p}^2.
    \end{equation}
    These identities are all formally derived in Appendix A.

    If the modulation strength $\alpha_\mathrm{e}$ is sufficiently weak, the optical response of the point-like dipole is governed by the fundamental and first negative Floquet harmonics. Hence, we can solve Eq.$\,$\ref{eq:drude-lorentz} in the frequency domain and write:
    \begin{equation}\label{eq:pol_sys}
        \begin{bmatrix}
        \vspace{2.0pt}
            \,\overline{\textbf{p}}_\mathrm{L}(\omega-\Omega)\, \\
            \overline{\textbf{p}}_\mathrm{L}(\omega)
        \end{bmatrix}
        = \hat\alpha(\omega)
        \begin{bmatrix}
        \vspace{2.0pt}
            \,\overline{\textbf{E}}_\mathrm{inc}(\omega-\Omega)\, \\
            \overline{\textbf{E}}_\mathrm{inc}(\omega)
        \end{bmatrix},
    \end{equation}
    where $\hat\alpha(\omega)$ is an effective polarizability tensor, given by:
    \begin{equation}\label{eq:pol_tensor}
        \hat \alpha (\omega)=\Delta
        \begin{bmatrix}
        \vspace{2.0pt}
            \frac{1}{\overline\alpha_\mathrm{L}(\omega-\Omega)} -\frac{\alpha_\mathrm{e}^2}{4} \tau  &
            \frac{\alpha_\mathrm{e}}{2}\lbrace \tau - \frac{1}{\overline\alpha_\mathrm{L}(\omega)}\rbrace  \\
            \frac{\alpha_\mathrm{e}}{2}\lbrace \tau - \frac{1}{\overline\alpha_\mathrm{L}(\omega-\Omega)}\rbrace  &
            \frac{1}{\overline\alpha_\mathrm{L}(\omega)} - \frac{\alpha_\mathrm{e}^2}{4} \tau 
        \end{bmatrix}^{-1}.
    \end{equation}
    Here, we have introduced:
    \begin{equation}
        \Delta = 1-\alpha_\mathrm{e}^2/4\, \text{, }\, \tau = \tilde\omega_\mathrm{rs}^2/(\varepsilon_0\,\tilde\omega_\mathrm{ps}^2)\,;
    \end{equation}
    and $\overline\alpha_\mathrm{L}(\omega)$ denotes the polarizability of the corresponding non-parametric Drude-Lorentz dipole,
    \begin{equation}
        \overline\alpha_\mathrm{L}(\omega_v) = \varepsilon_0\frac{\tilde\omega_\mathrm{ps}^2}{(\tilde\omega_\mathrm{rs}^2-\omega_v^2)-i \omega_v\tilde\gamma}.
    \end{equation}
    Eqs.$\;$\ref{eq:pol_sys} and \ref{eq:pol_tensor} are a key result of this paper. These equations enable 
    us to introduce an analytical model that provides deeper insight into the optical response of the time-modulated subwavelength particles by reducing them to a Drude-Lorentz point-like dipole. Its relevance is two-fold. On one hand, it allows us to find analytically the resonant modes of the time-varying sphere, unveiling their interplay and coalescence into an exceptional point (EP) with an analytical equation. On the other hand, it provides us with an  effective polarizability (Eq.$\,$\ref{eq:pol_tensor}). This is analogous to the effective polarizability usually defined in lattices of nanoparticles \cite{kravets-2018}, which characterize all the geometric effects of the array as well as the properties of the scatterers in a single quantity. In contrast, our effective polarizability, captures all frequency conversion and gain effects of the Floquet media, as well as the material and radiative properties of the nanoparticle, in a single quantity. Moreover, it allows us to evaluate the absorption cross-section by applying Eqs.$\,$31-32. 
    
\section{Unveiling the optical response with a two-band model analysis}

    \noindent We now utilize the two-band model to examine in detail the optical response of time-varying plasmonic nanoparticles obtained from exact T-matrix computations. First, we analyze the interaction between the two resonant modes as we shift the relevant parameters of the system: the modulation strength and the modulation frequency. 

    In Fig.$\;$3, we illustrate the optical response of the time-varying nanoparticle as a function of $\alpha_\mathrm{e}$ for a fixed value of modulation frequency, $\Omega=6.0$ eV (same as in Fig.$\,$2). In panel (a) we illustrate the particle's eigen-frequencies in the complex plane, derived from the Floquet T-matrix in the electrical dipole approximation. For each value of $\alpha_\mathrm{e}$, we show the two eigen-frequencies that correspond to the LSP and its amplifying counterpart, plotted with crosses of the same color. For low modulation strengths, starting with the dark blue cross at $\alpha_\mathrm{e}=0$, the real parts of the two eigen-frequencies are symmetric with respect to $\Omega/2$, while their imaginary parts are fixed at half the effective losses of the nanosphere. As $\alpha_\mathrm{e}$ increases, the interaction between the two resonances intensifies, and the real parts of their resonant frequencies drift towards each other until they coalesce at an EP, appearing for $\alpha_\mathrm{EP}=0.40895$ at $\mathcal{R}(\omega)=\Omega/2$. From here on, further increasing $\alpha_\mathrm{e}$ results in a shift of the imaginary part of the eigenmodes, while their real parts remains constant. Particularly, one of the eigen-frequencies shifts upwards along the imaginary axis, while the other drifts downwards acquiring enhanced loss. When the upward-shifting mode intersects the real axis, the nanoparticle reaches the parametric amplification condition ($\alpha_\mathrm{PA}=0.40965$). After this point, the eigen-frequencies' complex nature becomes the most significant component of the spectrum. 

    Figure 3(b) shows the absorption cross-section spectrum of the scatterer as a density plot, together with the mode resonant frequency derived from the two-band formalism, as a function of the modulation frequency. By looking at the particle's spectrum, we recognize the dynamics of the eigen-frequencies illustrated in panel (a). For low $\alpha_\mathrm{e}$, the spectrum exhibits a single maximum around the LSP. As the modulation strength grows, a negative-absorption dip emerges in the spectrum, and the conventional maximum redshifts while preserving the quality factor. In addition, the two local extrema are symmetric with respect to $\Omega/2$, and their positions match the real part of the eigenmodes of the particle discussed in panel (a). Notably, although both eigenmodes have eigen-frequencies with a negative imaginary part, their scattering properties differ because one of them is the replica of an eigenmode with negative-valued frequency, such that the LSP displays absorption and the replica displays gain. As the modulation strength approaches $\alpha_\mathrm{EP}$, the two local extrema converge at $\Omega/2$ and we identify a sudden increase in their absolute value, followed by the vanishing of both extrema in the spectra. The absorption enhancement is related to the parametric amplification  condition, that takes place when the system has a real-valued eigen-frequency ($\alpha_\mathrm{EP}\approx \alpha_\mathrm{PA}$), while the disappearance of the extrema results from the decoupling between imaginary-valued resonances and incident light with real-valued frequency. As a result, beyond $\alpha_\mathrm{PA}$ the influence of higher-order Mie eigenmodes becomes more pronounced in the absorption spectrum.

    In addition, in Fig.$\;$3(b) we show the eigenvalues of the system derived from the two-band model with an orange line on top of the absorption cross-section. For small $\alpha_\mathbf{e}$, the eigenmodes predicted by the two-band model closely reproduce the resonances of the system: the real parts of the two eigen-frequencies (solid orange line) nicely follow the absorption peak and gain dip. Specifically, when the modes are well separated in the frequency spectrum, their energy shift is well predicted by the simplified model and the resonant frequency reads: $\Delta\omega_\mathrm{lsp}$. Thus, our two-band model is able to accurately capture the frequency shifts from the LSP as a consequence of the temporal variation. Coalescence of the two eigenmodes in an EP is also nicely reproduced, albeit at a slightly lower value of $\alpha_\mathrm{e}$ than the one obtained from computations. The reason behind this is that the two-band model is a very good approximation for smaller modulations, while it is less accurate for higher $\alpha_\mathrm{e}$. After coalescing at the EP, the real part of the two eigenvalues stays at $\omega=\Omega/2$, and their two imaginary parts deviate from each other, in agreement with the discussion of Fig. 4(a). In addition, the two-band model provides analytical equations for the parameters $\alpha_\mathrm{e}$ at which the exceptional point and parametric amplification occur:
    \begin{subequations}\label{eq:al_ap_pa}
    \begin{align}
    \alpha_\mathrm{EP}^2 = & \,\frac{4}{\tilde\omega_{rs}^4}\lbrace \tilde\omega_{rs}^2 -(\Omega^2+\tilde\gamma^2)/4\rbrace^2,\\
    \alpha_\mathrm{PA}^2 = &\, \alpha_\mathrm{EP}^2 + 2 \frac{\tilde\gamma^2}{\tilde\omega_{rs}^4}\lbrace \tilde\omega_{rs}^2 + \Omega^2/4 \rbrace -\frac{\tilde\gamma^4}{4\tilde\omega_{rs}^4},
    \end{align}
    \end{subequations}
    for a system with fixed modulation frequency and radius. Eqs.$\,$\ref{eq:al_ap_pa} shows that in the limit $\tilde{\omega}_\mathrm{rs}\gg\tilde\gamma$, the condition for parametric amplification and the exceptional point occur at nearly the same modulation strength.  

    \begin{figure}[t]
        \includegraphics[width=0.95\columnwidth]{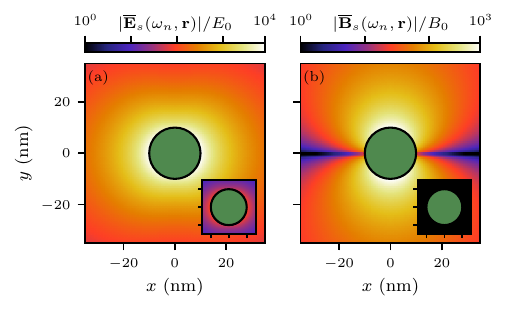}
        \caption{Near field enhancement by a periodically-modulated nanoparticle illuminated by a plane wave polarized along $x$ under the parametric amplification condition: $\omega=3.001$ eV, $\Omega=6$ eV and $\alpha_\mathrm{e}=$ $0.41$, with $R=10$ nm. The scattered electric (a) and magnetic (b) fields are obtained from Floquet T-matrix computations. The inset panels show the field enhancement in the absence of temporal modulation for comparison (incidence frequency $\omega=3.347$ eV).}
    \end{figure}

    \begin{figure*}[t]
        \includegraphics[width=\textwidth]{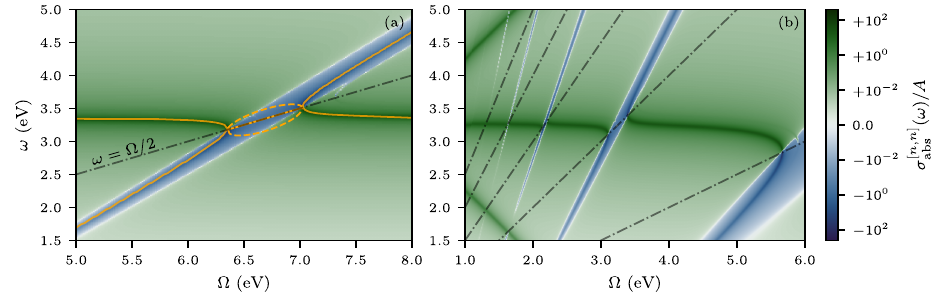}
        \caption{Absorption cross-section of a temporally modulated silver sphere with radius $R=10$ nm as a function of the modulation frequency $\Omega$ for fixed modulation strength: $\alpha_\mathrm{e}=0.2$ (a) and $\alpha_\mathrm{e}=0.6$ (b). The orange solid and dashed lines represent the real and imaginary parts of the system's eigenstates given by the two-band model, respectively. In addition, the doted-dashed gray lines represent $\omega=n\cdot\Omega/2$, with $n\in[1,5]$.}
    \end{figure*}

    We now discuss the field enhancement provided by the time-varying plasmonic nanoparticles in the parametric amplification regime. For this purpose, we compute from Eq.$\,$15 the fields scattered by a nanoparticle with a 10 nm radius and modulation frequency $\Omega=6.0$ eV (same as in Fig.$\,$3), under incident light at $\omega\sim\Omega/2$. We tune to the parametric amplification condition, $\alpha_\mathrm{e}=0.41$, and show in Fig.$\,$4 the spatial distribution of the scattered electric and magnetic fields. As the nanoparticles in the absence of modulation, shown in the inset panels for comparison, a time-varying subwavelength nanoparticle scatters with a dipolar radiation pattern. Moreover, gain provided by the modulation enables values of field enhancement two orders of magnitude larger than in the static case.


    Next, in Fig.$\;$5, we study the dependence on the modulation frequency in two regimes of modulation strength: $\alpha_\mathrm{e}=0.2$ (a) and $\alpha_\mathrm{e}=0.6$ (b). We show density plots of the normalized absorption cross section spectrum, computed through the T-matrix formalism as a function of the modulation frequency for a nanoparticle with radius $R=10$ nm. For the weaker modulation scenario, see Fig. 5(a), the spectrum features the interaction between the two modes discussed in Figs.$\;$2 and 3: the conventional LSP and its amplifying replica stemming from negative frequencies. In the non-interacting limit, the LSP yields a non-dispersive peak in the absorption spectrum at $\omega=\omega_{lsp}$, whereas the replica linearly disperses with the modulation frequency following $\omega=\Omega -\omega_{lsp} $, and results in a dip in absorption cross-section, seen in blue. Increasing $\Omega$, we observe that these features are overall maintained, but the interaction adds complexity when the two modes approach each other, as we can also see through the two-band model (eigenvalues plotted with an orange line). The two modes, symmetric with respect to $\omega=\Omega/2$, first approach each other as $\Omega$ increases. Then, as the modulation frequency approaches twice the LSP frequency, $\Omega=2\,\omega_{lsp}$, the two resonances coalesce. This feature is equivalent to the coalescence we discussed in Fig.$\;$3(a), and results in the formation of an $\Omega$-gap. Such a behavior is reminiscent of the momentum gaps in PTCs, and also originates from pairs of complex frequency states. In fact, the two eigenvalues share their real part across the band gap, while their imaginary parts split apart, one going towards negative values, the other one towards positive ones. This can also be seen with the dashed orange lines, which illustrate the imaginary parts of the eigenvalues as obtained from the two-band model. Further increasing $\Omega$, the two eigen-frequencies coalesce again into two real bands, and their real parts continue to follow the dispersions of both the LSP and its amplifying replica. Interestingly, since in this set of results $\alpha_\mathrm{e}$ is not too large, the simplified model is able to reproduce the behavior of the system accurately, even at the parametric amplification condition. In this regime, we can derive an explicit formula for the value of modulation frequency $\Omega$ at which parametric amplification arises, given a system with fixed modulation strength and radius: 
    \begin{equation}
        \frac{\Omega_{\pm}^2}{2} = (2\tilde\omega_{rs}^2-\tilde\gamma^2) \pm \lbrace\tilde\gamma^4-4\tilde\gamma^2\tilde\omega_{rs}^2+\alpha_\mathrm{e}^2\tilde\omega_{rs}^4\rbrace^\frac{1}{2},
    \end{equation}
    The two solutions, $\Omega_+$ and $\Omega_-$, define the boundaries of the $\Omega$-space gap, in which the real part of the eigenvalues equals half the modulation frequency. 

    \begin{figure*}[t]
        \includegraphics[width=0.95\textwidth]{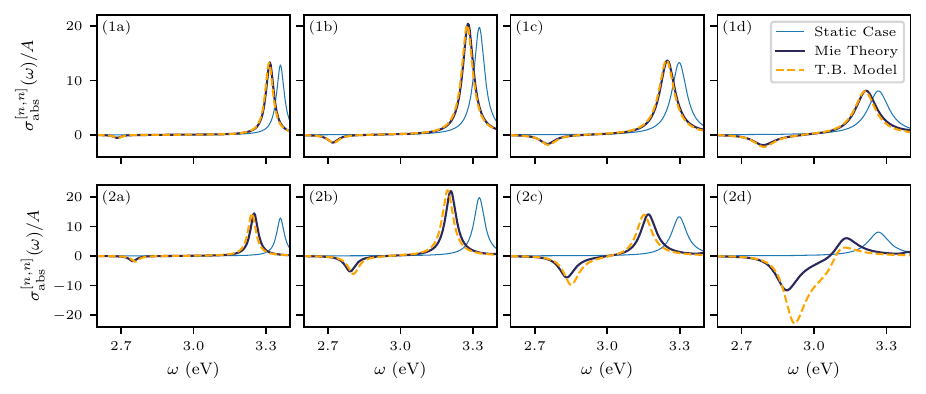}
        \caption{Normalized absorption cross-section of a time-modulated silver sphere with modulation frequency $\Omega=6.0$ eV in terms of the modulation strength: $\alpha_\mathrm{e}=0.2$ (1) and $0.3$ (2); and the particle size: $R=5$ nm (a), $15$ nm (b),  $20$ nm (c) and $25$ nm (d).}
    \end{figure*}

    Figure 5(b) shows the absorption cross-section spectrum for the scenario of stronger modulation, $\alpha_\mathrm{e}=0.6$. In this regime, we see the effect of higher-order Floquet replicas of the negative frequency partner of the LSP, which also result in amplifying windows, but for lower values of the modulation frequency. The phenomenology is similar to the one already discussed for the interaction between the LSP and the first-order amplifying Floquet replica, which can be seen around the coalescence at $\Omega\sim 5.5$ eV. Whenever $\omega_\mathrm{lsp}$ crosses with a high-order replica of order $n$, to which it is symmetric with respect to $\omega=n\Omega/2$ (with $n$ an integer number), the LSP and the replica coalesce at a half-integer of the modulation frequency, reaching new parametric amplification conditions, and resulting in new $\Omega$-space gaps with their associated amplification regime. Thus, in order to reach the parametric amplification regime, the modulation frequency may be reduced at the expense of larger modulation strengths. In addition, we note that the two-band model no longer describes accurately the dynamics of the new resonances. 

    Finally, we discuss the accuracy of the two-band model in more detail. Figure 6 shows the absorption cross section spectra of silver spheres derived using both methods provided in this work: the Floquet-Mie formalism (black lines) and the two-band model (dashed orange lines). We also plot the cross section of unmodulated nanoparticles for reference (thin blue lines). In specific, we seek to test the validity of the simplified model by exploring different values of sphere radius, which we change from $R=5$ nm in the first column to $R=25$ nm in the last column, and modulation strength. Starting with a smaller modulation strength, $\alpha_e=0.2$ (top row), we find that the two-band model accurately reproduces the absorption cross-section of nanospheres of radius up to $R=25$ nm. The simplified model is able to fully reproduce the redshift of the LSP compared to the unmodulated scenario, and its radiative broadening, which increases as the size of the nanosphere grows from panel (1a) to (1d). Additionally, it captures the emergence of the negative absorption dip, completely absent in the unmodulated case, and more prominent as the sphere's size increases. On the other hand, increasing the modulation strength to $\alpha_e=0.3$ (bottom row), while keeping small values of the radius, as in panels (2a) and (2b), with $R=5$ and 15 nm respectively, we observe that the two-band model is still able to accurately reproduce both the redshift of the LSP due to the time modulation, and the emergence of the negative absorption peak at the LSP replica. In contrast, further increasing the radius to $R=20$ and 25 nm, as in panels (2c) and (2d), we observe that the accuracy of the two-band model starts to fail. It anticipates a higher redshift of the LSP (and a higher blueshift of the replica) than the T-matrix's result, and more importantly, it overestimates the contribution of the amplifying mode, leading to much larger negative absorption peaks than the actual result. This is specially noticeable for the 25 nm case (2d). Finally, we note that the main reason behind the failing of the two-band model is its inability to reflect higher order Floquet interactions beyond the selected two-mode subspace, which becomes important as the modulation strength grows. In contrast, radiative effects, that become important within the larger nanoparticles, are accurately described, actually beyond the usual Meier-Wokaun polarizability for subwavelength particles \cite{meier}.

\section{\label{sec:level5}Conclusions}

    \noindent We have analyzed the optical properties of time-varying subwavelength metallic nanoparticles, utilizing both an exact scattering formalism based on the Floquet T-matrix, and a two-band simplified description based on a parametric Drude-Lorentz point-dipole model. We have shown that temporal modulation results in the appearance of a new resonant mode that exhibits a negative absorption cross-section and can interact with the conventional localized surface plasmon of the particle. For increasing strength of the temporal modulation, the interaction between these two modes increases, until their eigenfrequencies coalesce in an exceptional point and the system reaches parametric amplification, where losses are fully turned into gain. In this scenario, a large negative absorption cross section peak signals optical amplification, and is accompanied by very significant field enhancements in the vicinity of the plasmonic nanoparticle. On the other hand, the recently introduced pseudounitary Floquet scattering-matrix \cite{globosits-2024} could be applied to our system in order to characterize the emergence of lasing and coherent perfect absorption \cite{globosits-2025}.

    Furthermore, we have proposed a simplified analytical description based on describing a periodically-modulated subwavelength plasmonic particle as a parametric Drude-Lorentz point-dipole. We have shown that the two mode approximation of this model is accurate for a wide range of parameters. Moreover, we have introduced an effective two-frequency polarizability, that allows us to obtain an analytical description of the system, providing insight on the optical response of time-varying scatterers. Our two-band description identifies the modes responsible for the plasmonic and amplifying resonant modes supported by the modulated particle, and unveils how their interaction leads to the optical spectra of the time-varying particle, including eigenfrequencies shift and coalescence. Finally, we note that our point-dipole model and analytical effective polarizability can be useful for studying the properties of arrays made of time-varying plasmonic nanoparticles, and can shed light on fully numerical approaches in order to study the optical response from time-dependent scatterers \cite{garg-2022}. We foresee that the interplay between time-varying media and plasmonics will provide a rich platform for nanoscale optics, opening up the possibility of realizing optical am- plification without the need of gain media.

\begin{acknowledgments}

    \noindent We acknowledge financial support from the EU (ERC 
     grant TIMELIGHT, GA 101115792) and MCIUN/AEI (PID2022-141036NA-I00 through MCIUN/AEI/10.1303 9/501100011033 and FSE+; RYC2021-031568-I; Programme for Units of Excellence in R\& D CEX2023-001316-M).
    
\end{acknowledgments}

\appendix

\section{Drude-Lorentz Description For Unmodulated Nanoparticles}

    \noindent Within the Floquet-Mie T-matrix formalism, the scattering cross-section of a subwavelength plasmonic sphere in the absence of temporal modulation is approximated by:
    \begin{equation}
        \sigma_\mathrm{sca}^{[n,n]}(\omega)
        \approx\frac{6\pi}{k_n^2}\left|\lbrace {\hat{T}^\mathrm{sca}_{N,1}(\omega)}\rbrace_n^n\right|^2
        = \frac{6\pi}{k_n^2}\left| \frac{R_n}{R_n+iI_n} \right|^2,
    \end{equation}
    where we use the following definitions:
    \begin{gather}
        R_n= \tilde{n}^2j(\tilde{n}x_n)\left[x_nj(x_n)\right]'-j(x_n)\left[\tilde{n}x_nj(\tilde{n}x_n)\right]',\\
        I_n=\tilde{n}^2y(\tilde{n}x_n)\left[x_ny(x_n)\right]'-y(x_n)\left[\tilde{n}x_ny(\tilde{n}x_n)\right]',
    \end{gather}
    with $j(x)$ and $y(x)$ denoting the 1st order spherical Bessel and Neumann functions, respectively. These expressions can be expanded as power series in $x_n$, providing explicit analytic equations valid in the subwavelength regime \cite{fan-2014}:
    \begin{gather}
        R_n\approx \frac{2}{9} x_n^2 \sqrt{\varepsilon_n}\left[\varepsilon_n-1\right], \\
        I_n\approx \frac{1}{3x_n}\sqrt{\varepsilon_n}
        \left[ \varepsilon_n + 2 - \frac{x_n^2}{2}\left( \varepsilon_n-1 \right)\left( \frac{\varepsilon_n}{5}+2 \right) \right],
    \end{gather}
    with $\varepsilon_n=\varepsilon_\mathrm{s}(\omega_n)$ denoting the static relative permittivity satisfying
    \begin{equation}
        \Re[\varepsilon_n]\approx\varepsilon_\infty-\omega_\mathrm{p}^2/\omega_n^2 \gg\Im[\varepsilon_n]\approx\gamma\,\omega_\mathrm{p}^2/\omega_n^3.
    \end{equation} 
    These approximations lead to a simplified expression for the scattering matrix element:
    \begin{equation}
        \lbrace\hat{T}^\mathrm{sca}_{N,1}(\omega)\rbrace_n^n \approx
        \frac{\frac{2}{3}x_n^3(\Re[\varepsilon_n]-1)}{\frac{2}{3}x_n^3(1-\Re[\varepsilon_n])-\Im[\varepsilon_n]+ i g(\omega_n)} ,
    \end{equation}
    where $g(\omega)$ is given by:
    \begin{equation}
        g(\omega_n) = \Re[\varepsilon_n] + 2 -\frac{x_n^2}{2}(\Re[\varepsilon_n]-1)(\Re[\varepsilon_n]/5+2),
    \end{equation}
    and cancel out at the localized surface plasmon resonance frequency of the sphere, $\omega_\mathrm{lsp}$:
    \begin{equation}
        g(\omega_\mathrm{lsp})=0\Rightarrow \Re[\varepsilon_n]=-2-\frac{12}{5}x_n^2 + \mathcal{O}(x_n^4).
    \end{equation}
    Hence, the resonance of the particle depends on both its size and the quasistatic limit resonance $\omega_q$,      
    \begin{equation}
        \omega_\mathrm{lsp}^2(R)=\frac{
        -5c^2\omega_\mathrm{p}^2 + \sqrt{25c^4\omega_\mathrm{p}^4+240c^2R^2\omega_\mathrm{p}^2\omega_\mathrm{q}^4}
         }{24R^2\omega_\mathrm{q}^2}.
    \end{equation}
    Finally, around the localized surface plasmon resonance, the following relation holds
    \begin{equation}
        g(\omega_n) \approx g'(\omega_\mathrm{lsp}^2)(\omega_\mathrm{lsp}^2-\omega_n^2)=\frac{1}{\beta_n}\frac{\omega_\mathrm{p}^2}{\omega_\mathrm{lsp}^2}(\omega_\mathrm{lsp}^2-\omega_n^2),
    \end{equation}
    where $\beta_n=\lbrace1-9x_n^2/10\rbrace^{-1}$, which enables us to express the scattering cross-section of the nanoparticle as
    \begin{equation}
        \sigma_\mathrm{sca}^{[n,n]}(\omega)
        \approx \frac{6\pi}{k_n^2}
        \left|\frac
        {2x_n^3\omega_\mathrm{lsp}^4/\omega_\mathrm{p}^2}
        {(\omega_\mathrm{lsp}^2-\omega_n^2)+i\omega_n (\gamma+\gamma_\mathrm{rad})} 
        \right|^2,
    \end{equation}
    with $\gamma_\mathrm{rad}=2R^3\omega_\mathrm{lsp}^6/(c^3\omega_\mathrm{p}^2)$ being the radiative losses of the nanoparticle.

    In contrast, the cross-section of a Drude-Lorentz point-like dipole with time-independent parameters is given by:
    \begin{multline}
        \sigma_{sca,p}^{[n,n]}(\omega)
        = \frac{k_n^4}{6\pi\varepsilon_0^2} \left|\,\overline\alpha_\mathrm{L}(\omega_n)\right|^2 \\
        = \frac{k_n^4}{6\pi}\left|\frac{\tilde\omega_\mathrm{ps}^2}{(\tilde\omega_{rs}^2-\omega_n^2)-i \omega_v\tilde\gamma}\right|^2,
    \end{multline}
    where $\tilde\omega_\mathrm{ps}$ stands for the effective static plasma frequency, $\tilde\omega_\mathrm{ps}$ represents the resonant static frequency and $\tilde\gamma$ is the effective damping factor. Comparing the scattering cross sections of the static sphere and the corresponding dipole, we derive the following identities:
    \begin{gather}
        \tilde\omega_\mathrm{rs}^2 = \omega_\mathrm{lsp}^2, \\ 
        \tilde\omega_\mathrm{ps}^2=12\pi R^3\omega_\mathrm{lsp}^4/\omega_\mathrm{p}^2,\\
        \tilde\gamma = \gamma + \gamma_\mathrm{rad}.
    \end{gather}
    Notably, in the quasistatic limit both effective frequencies are related to the charge-carrier density: $\tilde\omega_\mathrm{rs:q}^2,\tilde\omega_\mathrm{ps:q}^2\propto N$. Hence, modulating in time the latter parameter induces similar temporal dynamics in both the effective resonant and plasma frequencies. To derive the two-band descrip- tion introduced in Sec.$\;$IV we assume that this temporal dependence holds in the subwavelength regime.

    \newpage

\bibliography{apssamp}

\end{document}